\documentclass[prb,twocolumn,showpacs,superscriptaddress,preprintnumbers,amssymb]{revtex4}
\usepackage{graphicx}
\usepackage{dcolumn}
\usepackage{bm}
\usepackage{amssymb}
\usepackage{amsfonts}

\newcommand{\comment}[1]{}

\newcommand{\beq}{\begin{equation}}
\newcommand{\eeq}{\end{equation}}
\newcommand{\beqn}{\begin{eqnarray}}
\newcommand{\eeqn}{\end{eqnarray}}

\begin{document}

\title{Coherent phonon scattering effects on thermal transport in thin semiconductor nanowires}
\author{P.~G.~Murphy}
\affiliation{Department of Physics, University of California,
Berkeley, CA 94720}
\author{J.~E.~Moore}
\affiliation{Department of Physics, University of California,
Berkeley, CA 94720} \affiliation{Materials Sciences Division,
Lawrence Berkeley National Laboratory, Berkeley, CA 94720}
\date{\today}
\begin{abstract}
The thermal conductance by phonons of a quasi-one-dimensional solid with isotope or defect scattering is studied using the Landauer formalism for thermal transport.  The conductance shows a crossover from localized to Ohmic behavior, just as for electrons, but the nature of this crossover is modified by delocalization of phonons at low frequency.  A scalable numerical transfer-matrix technique is developed and applied to model quasi-one-dimensional systems in order to confirm simple analytic predictions.  We argue that existing thermal conductivity data on semiconductor nanowires, showing an unexpected linear dependence, can be understood through a model that combines incoherent surface scattering for short-wavelength phonons with nearly ballistic long-wavelength phonons.  It is also found that even when strong phonon localization effects would be observed if defects are distributed throughout the wire, localization effects are much weaker when defects are localized at the boundary, as in current experiments.
\end{abstract}
\maketitle
\def\gmn{g_{\mu\nu}}
\def\beq{\begin{equation}}
\def\eeq{\end{equation}}

\section{Introduction}
\label{sec:intro}

The transport of heat by phonons in quasi-one-dimensional structures is sensitive 
to quantum confinement effects once the phonon wavelengths that dominate thermal 
transport are comparable to the structure dimensions.  Strong quantum 
confinement in thermal transport has been demonstrated in the 
observation~\cite{schwab} of 
the thermal conductance quantum~\cite{rego,cross, tanaka} 
at temperatures T $\leq 1$ K and 
also in recent experiments on semiconductor nanowires~\cite{li,lisuper} and 
nanotubes~\cite{mceuentherm} at higher temperatures (down to $\sim$ 10 K).  

\begin{figure}
\includegraphics[width=3in,height=2.2in]{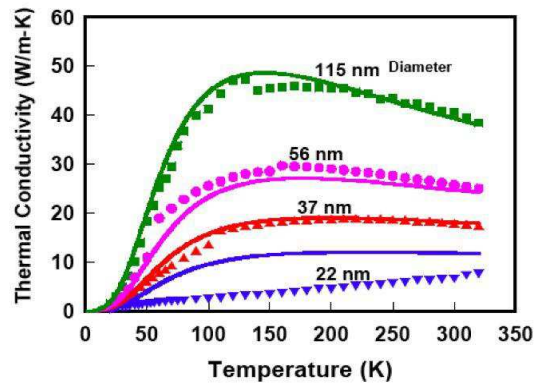}
\caption{
The thermal conductance as a function of temperature
for various wire diameters. 
The fit (solid lines) matches the data to a Boltzmann equation
treatment including impurity, surface and umklapp scattering.
Note that the 22nm wire is anomalous.
This data is taken from Li et al., Appl.~Phys.~Lett, {\bf 83} 2934 (2003) 
(Ref.~\onlinecite{li}); the fit (except for $d=22$) is from N. Mingo, Phys.~Rev.~B, 
{\bf 68} 113308 (2003)
(Ref.~\onlinecite{mingo}); figure courtesy A.~Majumdar. 
} 
\label{li_mingo}
\end{figure}

Experiments on the thinnest nanowires are not well explained by current models of 
independent scattering events at the boundary~\cite{mingo}, even though such models describe thicker wires quite well (see figure \ref{li_mingo}).  In particular, data on a 22\ nm Si nanowire~\cite{li} shows thermal conductivity scaling with temperature up to temperatures of order 200 K, well above the temperature where higher transverse modes should be occupied; linear temperature dependence~\cite{schwab} is expected when only the four gapless modes are occupied.  This paper studies the effect of 
coherence between different scattering events in quasi-one-dimensional systems of 
variable transverse dimension, using a large-scale numerical transfer matrix 
approach to check simple analytic models.  The differences between bulk and boundary disorder are studied \cite{froufe, blencowe_phonon_localization, kambili}, and linear temperature dependence at high temperature is found to depend on having boundary disorder rather than bulk disorder.

The study of quantum effects on electrons in quasi-one-dimensional systems has developed steadily since the observation in 1988 of quantized electrical conductance in ballistic quantum point contacts~\cite{vanwees}.  The electrical conductance as a function of gate voltage shows plateaus at multiples of $G_e = 2 e^2 / h$, where $h$ is Planck's constant and $e$ the electron charge, and the factor of 2 results from spin.  Hence $G_e$ is a universal conductance quantum for fermions and insensitive to material properties.  The corresponding thermal conductance quantum has recently been measured in low-temperature heat transport through small insulating structures~\cite{schwab}: at the lowest energies the only modes accessible in this system are four phonon modes, and the observed thermal conductance at the lowest temperatures is
\beq
G = 4 g_0 = 4 {\pi^2 k_B^2 T \over 3 h},
\label{thermal_quantum}
\eeq where $T$ is temperature and $k_B$ the Boltzmann constant.


This paper studies how scattering and localization by disorder appear in the thermal conductance of long quasi-one-dimensional structures.  Previous work on coherent scattering in phonon transport has concentated on the strictly one-dimensional limit~\cite{akkermans}; as explained below, there are fundamental aspects of phonon confinement even in the clean case that are not well captured by a purely 1D chain.  We find several important differerences between scattering and localization effects in 1D phonon transport and the corresponding effects in electrical transport.  As in the electronic problem, transport in quasi-1D systems can be ballistic, diffusive, or localized, but unlike the electronic problem, the lowest four harmonic modes are protected from scattering as $T\rightarrow 0$ by fundamental properties of rigid bodies (in the same way as the three acoustic modes of bulk solids).  In order to access very large system sizes, many of our numerical calculations are carried out on in two rather than three spatial dimensions; however, two-dimensional systems already include some important features absent in one dimension, e.g., different dispersion relations for the lowest-lying modes ($\omega \sim k$ for torsional and longitudinal, $\omega \sim k^2$ for transverse).
This paper presents two separate approaches to understanding the thermal conductance
of these thin nanorods. Firstly, we use a scalar model to study the mean free path
of different phonon modes. We find that the gapless mode (which has no transverse 
wavevector) has a mean free path significantly longer than the length of the system;
when combined with the contribution of the other modes, the resulting
thermal conductance for very thin rods (of order 20 nm) is approximately linear.

We also obtain the thermal conductance at low temperature by numerical transfer-matrix calculations on systems of small transverse dimension and an analytic theory 
that makes standard mesoscopic assumptions about the nature of scattering.  
Calculations are done using the Landauer formalism valid at low 
temperatures~\cite{angelescu,rego,blencowe,fagas}.  
The analytic results for boundary scattering build on the 
scalar approximation introduced in Ref.~\onlinecite{santamore_prb_scalar}.  The 
coherent multiple scattering discussed in this paper is much more important in 
long nanowires and nanotubes, where the ratio of length to width may be 100:1, than 
in the suspended membrane devices studied in Ref.~\onlinecite{schwab}, where 
nonuniform width and individual scattering effects have been shown to explain 
the experimental data~\cite{cross,blencowe}. 

The results obtained here suggest that the experimentally observed dip in $G/T$ at moderate temperatures, discussed in detail for the experimental geometry of Ref.~\onlinecite{schwab} in Ref.~\onlinecite{cross}, will appear generically in one-dimensional systems as a consequence of delocalization of phonons at low energy and 1D-3D dimensional crossover at high energy.  Our theoretical predictions are also compared to experiments on semiconductor nanowires at relatively high temperature as a function of diameter~\cite{li}.  We find that, while the linear temperature dependence on the smallest nanowire in this experiment is a robust phenomenon, the coefficient does not need to be close to $4 g_0$, although it happens to be remarkably close to this value in that experiment.

We use the term "phonon" to describe both long-wavelength bulk modes of the nanowire and conventional short-wavelength phonons as there is no sharp distinction between these two limits.  Note that phonon localization by strong local scattering can occur also in $d>1$ (a review is Ref.~\onlinecite{allen}), but considerably more disorder is required. 
The following section discusses the basic theory of mesoscopic thermal transport in harmonic quasi-one-dimensional systems: localization is described by the DMPK model with appropriate modifications.  In section III we argue for the relevance of Anderson localization
and mode inequivalence for thin nanowires.  Section IV contains numerical studies of a 2D model to understand the key differences between phonon and electron mesoscopics and test the DMPK model, and Section V applies this model to three-dimensional nanowires at low temperature.

\section{Mesoscopic thermal transport}
\label{sec:meso}


Our approach is based on the thermal analogue of the Landauer formalism for electronic transport.  The net thermal current in a wire from a reservoir at temperature $T+\delta T$ to one at temperature $T$ is the difference between that for the phonons travelling to the right minus that for the phonons travelling
to the left:
\beqn
J &=& 
\sum_i \int {dk\over 2\pi}\hbar\omega_i(k) v(k)
{1\over e^{\hbar\omega(k)/ k_B (T+\delta T)} -1}{\cal T}_i (\omega)\cr&&
- \sum_i \int {dk\over 2\pi}\hbar\omega_i (k) v(k)
{1\over e^{\hbar\omega_i (k)/ k_B T} -1}{\cal T}_i (\omega).
\eeqn
Here the summation index $i$ runs over all propagating phonon modes and ${\cal T}_i(\omega)$ is a transmission probability of phonons in mode $i$ at frequency $\omega$, defined more precisely in terms of the heat flux below.  The boundary condition is that a possibly disordered segment of finite length is connected at both ends to infinite reservoirs with no disorder.  Note that the distribution in the center of the wire is {\it not} a thermal distribution, and has no well-defined temperature: the Landauer approach hence does not apply if dissipation within the wire leads to thermalization~\cite{nose,hoover}.  The wire is assumed to be perfectly harmonic unless otherwise stated; reduction of thermal transport by anharmonic scattering is exponentially damped below the Debye temperature.~\cite{ashcroft}

The thermal conductance $G$ of a nanowire with perfect transmission (${\cal T}_i=1$) of harmonic vibrational modes from a thermal distribution at temperature $T$ is $G = J/\delta T$, or
\beq
G = {1 \over 2 \pi \hbar} \int_0^\infty N(\omega) {\hbar^3 \omega^2 \over k_B T^2} {e^{\hbar \omega / k_B T} \over (e^{\hbar \omega / k_B T} - 1)^2}\,d\omega.
\label{landauer}
\eeq
Here $N(\omega)$ is the number of propagating modes at frequency $\omega$ and $k_B$ is Boltzmann's constant.  For a 3D elastic rod, such as a semiconductor nanowire, there are four modes that survive down to zero frequency: one torsional and one longitudinal mode, each with linear dispersion $\omega \sim k$, and two flexural models with quadratic dispersion $\omega \sim k^2$.

Note that this conductance is finite even if there is perfect transmission of heat from one end of the wire to another: this is the thermal equivalent of the well known ``contact resistance'' in 1D electronic systems.  If $N(\omega)$ is constant, so that $N(\omega) = N(\omega=0)$,
we obtain in this ballistic limit
\beq
G = N(0) {\pi^2 {k_B}^2 T \over 3 h}.
\eeq
The same formula gives the leading behavior for small temperature $T$ even if $N(\omega)$ is not constant, because in this limit the Bose-Einstein factor in the integral becomes concentrated near $\omega = 0$.

Elastic scattering is included in the Landauer approach through the ``throughput'' ${\cal T}(\omega) = \sum_i {\cal T}_i(\omega)$, which gives the fraction of incident energy flux on the left end that is transmitted through to the right end (hence ${\cal T}(\omega) = N(\omega)$, the number of modes,  for clean systems with no scattering).  The conductance is then
\beq
\label{therm_cond}
G = {1 \over 2 \pi \hbar} \int_0^\infty {\cal T}(\omega) {\hbar^3 \omega^2 \over k_B T^2} {e^{\hbar \omega / k_B T} \over (e^{\hbar \omega / k_B T} - 1)^2}\,d\omega.
\eeq

The numbers ${\cal T}_i(\omega)$ for disordered 1D systems become exponentially small once the length of the 1D wire is larger than a ``localization length'' $\xi$ that is determined by disorder strength and wire geometry.  The analytic model of Maynard and Akkermans ~\cite{akkermans} on a pure 1D chain ($N(\omega) = 1$ for all allowed $\omega$) with random masses can be summarized as follows: assume the throughput ${\cal T}(\omega)$ is unity for $L < \xi$ and zero for $L>\xi$, where the localization length $\xi$ for the 1D linear chain follows from results of Dyson~\cite{itzykson}:
\beq
\xi(\omega)= {8 {\omega_D}^2 \over {\sigma_M}^2 \pi^2 \omega^2}.
\label{dyson}
\eeq
Here ${\sigma_M}^2$ is the variance of the random mass distribution and $\omega_D$ is the Debye frequency.  Note that the frequency dependence of $\xi$ means that the same wire can be in the localized regime (i.e., $L \gg \xi$) for high frequencies, while in the ballistic regime (i.e., $L\ll\xi$) for low frequencies.

There are three regimes in a quasi-one-dimensional (multimode) wire. Let $\ell(\omega)$ be the mean free path averaged over the modes that propagate with frequency $\omega$.
At a given $\omega$, if $N\ell \gg L$ then
\beq
{\cal T}(\omega) = 
{N(\omega) \over 1+L/\ell(\omega)} + {\cal O}(N^0)
\label{ohmic}
\eeq
This equation describes ballistic behavior for $\ell(\omega) \gg L$, and 
for $N \ell \gg L \gg \ell$, Ohmic behavior in which the thermal conductance of a long wire decreases as 
the reciprocal of the length.  If $N \ell \ll L$, then the phonons at frequency 
$\omega$ are effectively localized and
\beq
{\cal T} \sim e^ {-L /\xi}
\label{local}
\eeq
where the localization length is $\xi = N\ell$.  The number of propagating modes $N$ clearly plays an important role in determining the crossover from Ohmic to localized transport.


These conclusions follow from noting that the same DMPK equation~\cite{dorokhov,mpk,beenakker} approach valid for electronic systems is appropriate for phonons, except that the dependence of the mean free path $l$ on frequency is modified.  The connection between ${\cal T}$ and the observed transport coefficient, expressed in the Landauer formula, is also different for electrons and phonons, since in the phonon case all frequencies with $\hbar \omega \leq k_B T$ contribute significantly, while in the electronic case only frequencies within $k_B T$ of the Fermi level contribute significantly.

The DMPK equation is valid as long as phonon transfer matrices on scales longer than some mean free path become described by the orthogonal random matrix ensemble for one-dimensional systems~\cite{beenakker}.  While we cannot prove the validity of random-matrix theory assumptions for this problem, Section \ref{sec:numerics}  verifies that the Ohmic (\ref{ohmic}) and localized (\ref{local}) limits for ${\cal T}$ correctly describe thermal transport in a model system with {\it bulk} impurity scattering, while strongly localized transport in models with {\it boundary} scattering is not observed.  Section \ref{sec:transfer_matrix} explains how microscopic transmission coefficients are calculated via a transfer-matrix approach. In the next section we analyze the case of surface scattering.

\section{Strongly disordered surface}
\label{sec:loc}

It is typically assumed that the mean free path of a phonon due to surface
scattering is or the order of $d$, the width of the material. In this
section we show that this is not true for all phonon modes.
Consider, for instance, the gapless modes (ie. those that have vanishing
wave-vector perpendicular to the long direction of the rod.) 
It is possible to estimate the mean free path of these modes using the
following model.
Consider a thin three-dimensional rod of length $L$, of square cross-section with
each edge of width $d$.
Allow scalar waves to propagate on the strip, with Neumann boundary
conditions. The dispersion relation is then
\beq
\omega = c \sqrt{ k^2 + k_\perp^2} 
\eeq
where $k_\perp = \sqrt{n_y^2 + n_z^2 }(\pi/d)$, and $n_x, n_y=0, 1, 2, \ldots$. We seek to
calculate the mean free path due to surface scattering of the $n_x=0,n_y=0$ mode.

It can be shown using
the methods of Ref.~\onlinecite{santamore_prb_scalar} that, for 
this mode, 
the mean free path is given by
\beq
\label{eqn:l0}
\ell_0^{-1} (\omega) =  
4 (3\pi^2)^{2/3} {h^2 \over d^3} \left({\omega \over\omega_D}\right)^2
 + 6\pi (3\pi^2)^{1/3} { (h/a)^2 \over d} \left({\omega \over\omega_D}\right)^4
\eeq
where $h$ is the mean width of the disordered surface layer.
$c$ is the velocity of the gapless mode, and the correlation
length scale for the surface disorder has been set to $a$, the lattice spacing. 
The first term comes from scattering back into the gapless mode, and the second
into all other modes, and so involves a factor of the density of states, which scales
like $\omega^2$.
The frequency dependence here is
what one expects for scattering from point-like impurities. 
The gapless mode, then, does not sense that the disorder is at the boundary: it
sees the disorder as a collection of point-like impurities.
This frequency dependence ensures that at low temperatures the
mode will be scattered weakly, and so will ``reflect'' almost specularly
from the boundary. 

This weak scattering of the gapless mode
agrees with Rayleigh's criterion on the condition for 
specular reflection from a boundary.
Rayleigh considered the following simplified model of a disordered surface:
suppose the clean surface is a straight line at $y=0$. 
Let the surface be at either $y= h/2$ or $y=-h/2$. An
incident wave then reflects from the disordered boundary
as from two Bragg planes, a distance $h$ apart. Define $\theta$
as the angle the incident wave makes with the surface.
The phase difference
between different parts of the reflected wave is then given by
\beq
\Delta \phi = 2\pi {2 h \sin\theta \over \lambda} = 2 h k_\perp
\eeq
where $k_\perp$ is the component of the incoming wavevector perpendicular to
the surface. If this phase difference $\Delta\phi$ is of order $1$, then the reflected
wave is destroyed through destructive interference, and the so wave must have been
reflected diffusely. On the other hand, if $\Delta \phi \ll 1$, the
reflection will be specular.

Consider now the phonon modes of the quasi-one-dimensional rod we considered above. 
One expects then that the modes with lower $k_\perp$
will scatter more weakly from the boundary than those with a larger value.
As we increase $k_\perp$ through the value $1/h$, one expects a transition
from modes with a long, frequency-dependent mean free path to modes with a 
mean free path of order $d$. 

\begin{figure}
\includegraphics[width=3in,height=2.2in]{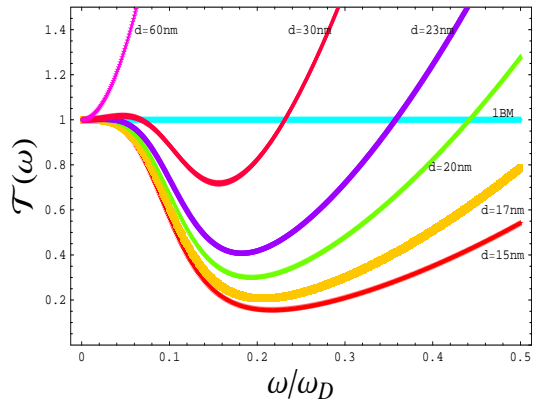}
\caption{ The throughput as a function of frequency for a surface
disordered rod, as suggested by equation (\ref{eqn:ansatz}). The length was
chosen to be 2000 nm, a typical length for a nanowire. Here
we have assumed that only the gapless mode has a frequency dependent
mean free path, and all others have a mean free path of order the
width of the system, $d$. The label ``1BM'' stands for ``one ballistic mode''.
}
\label{tp_analytic}
\end{figure}

\begin{figure}
\includegraphics[width=3in,height=2.2in]{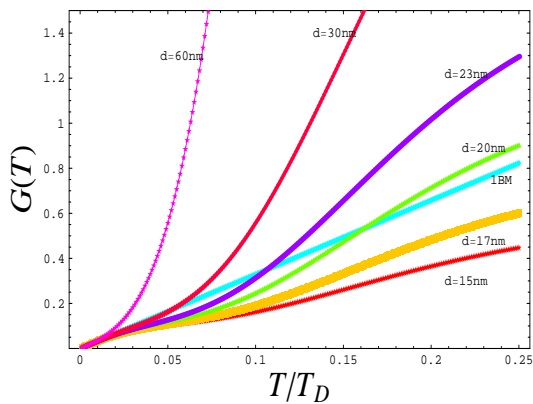}
\caption{ The thermal conductance associated with the throughput functions
shown in figure \ref{tp_analytic}. The scale for $G(T)$ is set by the
curve for one ballistic mode (1BM). We have also included the correction due
to umklapp scattering using the form suggested by Mingo \cite{mingo}, but this
correction is very small in the temperature range of interest.
}
\label{thermal_cond_num_int}
\end{figure}

We can make an ansatz for the total throughput in the case of strong 
mode inequivalence, assuming that only the gapless mode has the frequency
dependent mean free path, and all others have a mean free path of $d$:
\beq
\label{eqn:ansatz}
{\cal T} (\omega) = \sum_{i} {1 \over 1+ L/\ell_i}
\eeq
where $\ell_i = \ell_0$ for the gapless mode and equals $d$ for
all other modes.
This formula gives the DMPK form in the case of mode equivalence 
($\ell_i = \ell$), and gives ballistic behavior if one of the 
mean free paths is much longer than the length of the system.
We have assumed that the strongly scattered modes give an
Ohmic contribution. In detail then, the throughput for the rod is
\beq
{\cal T} (\omega) = {1 \over 1+ L \ell_0^{-1}} + {d \over L} {(3\pi^2)^{2/3} \over 8\pi}
{d^2 \over a^2} \left({\omega\over\omega_D}\right)^2
\eeq
This is plotted for various widths in figure \ref{tp_analytic}. This function can also
be numerically integrated using equation \ref{therm_cond} 
to find the corresponding thermal conductance
as a function of frequency, $G(T)$. This is shown in figure \ref{thermal_cond_num_int}.

The throughput is the sum of two contributions: the gapless mode gives
a term that scales like $1/(1+ C \omega^4)$; the other modes a term 
that scales like $\omega^2$. Together, for a sufficiently thin rod, 
the dip in the throughput is somewhat smeared out by the exponential
factors in the integral for $G(T)$, with the result that
the thermal conductance is approximately
linear in $T$ up to about $T_D/5$.

It is also interesting to consider whether 
localization effects might be relevant.
Let us assume that, at a frequency $\omega$, there are of order $N(\omega)$ modes 
with a mean free path of $d$ (we ignore for the moment the quasi-specular modes).
For this given mean free path, the localization length becomes
$\xi(\omega) \simeq  N(\omega) d$. For low frequencies, the number
of modes is small, and so $\xi$ should be of order the width of the
system, which by assumption is much
shorter than the length $L$. Hence all the low frequency modes should be localized.
As we increase the frequency, the number of modes increases like 
\beq
N(\omega) \sim \left( {\omega\over v/d}\right)^2
\sim \left( {\omega\over \omega_D}\right)^2\left({d\over a}\right)^2
\eeq
and so at some value $\omega_*$ we will have $ N(\omega_*) \alpha d = L$, or 
\beq
\omega_* \sim \omega_D \sqrt{L/a \over (d/a)^3} \sim 0.24 \omega_D 
\left( {22nm\over d}
\right)^{3/2}
\eeq
where we've set $a = 0.543$ nm, the lattice constant of silicon, $L = 2$ $\mu$m, 
a typical value for the length of a nanowire, and 22 nm was chosen as 
a comparison scale, since this is the width of the wire with anomalous
thermal conductivity as observed by Li et al \cite{li}.
The frequency scale $\omega_*$  corresponds to a temperature scale of 
\beq
T_* \sim 0.24 T_D \left( {22nm\over d}\right)^{3/2}
\eeq
Under the assumption then these phonons have a mean free path due
to boundary scattering of order $d$, one should observe an exponentially small
contribution to thermal transport at temperatures less than $T_*$.

\begin{figure}
\includegraphics[width=3in,height=2.2in]{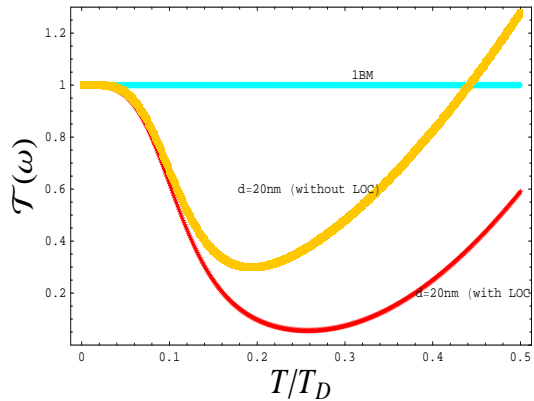}
\caption{ The throughput as a function of frequency for a surface-disordered rod, 
to illustrate the relevance of localization.
based on the formula \ref{eqn:tploc} in the text.
As before, ``1BM'' stands for ``one ballistic mode''; the ``with loc'' curve
is based on the formula \ref{eqn:tploc} in the text, and the ``without loc''
curve is identical to that of figure \ref{tp_analytic}.
}
\label{tp_analytic_loc}
\end{figure}

\begin{figure}
\includegraphics[width=3in,height=2.2in]{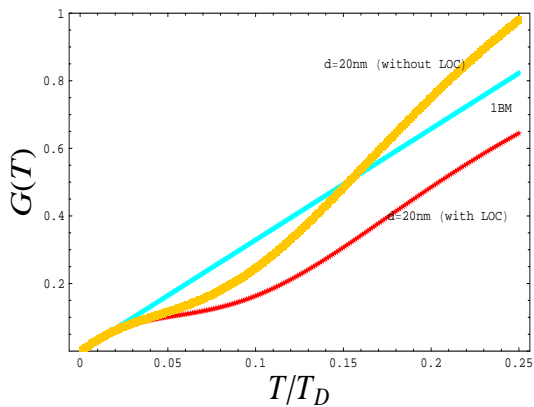}
\caption{ The thermal conductance associated with the throughput functions
shown in figure \ref{tp_analytic_loc}. Again, the scale for $G(T)$ here is set by the 
one ballistic mode (1BM).
}
\label{thermal_cond_num_int_loc}
\end{figure}

These considerations are of course greatly complicated by the mixing with the 
quasi-specular modes. The DMPK theory assumes that the transmitting channels
are equivalent, and it is not known how the localization length changes
when a quasi-ballistic mode is added to a number of strongly scattered modes.
As an ansatz we approximate the quasi-ballistic  mode as having decoupled
from the strongly scattered modes. The resulting throughput is
\beq
\label{eqn:tploc}
{\cal T} (\omega) = {1\over 1+ L/\ell_0}+  {N(\omega) \over 1+ L/d} e^{-L/(N(\omega)d)}
\eeq
The throughput and corresponding thermal conductances are shown in 
figures \ref{tp_analytic_loc} and \ref{thermal_cond_num_int_loc}. 
Localization significantly reduces the thermal conductance below the 
incoherent limit.

A full understanding would require a theory that incorporates
inequivalent channels, something that is not available at this time.
Our numerical work described later sheds much light on this problem.

Note that in the high-frequency limit where it is possible to make a wave packet 
of phonons much smaller than the width of the wire, and this wave packet scatters 
like a particle, boundary scattering allows wave packets directed along the rod 
axis to pass through: this implies that true exponential length dependence is not 
expected with boundary scattering, even when most thermal phonon modes are localized.

\section{Dynamical transfer matrix approach}
\label{sec:transfer_matrix}

First recall the definition of scattering and transfer matrices 
for waves.
Let the amplitude in mode $i$ that is incoming from the left be $a_i$; incoming from 
the right be $d_i$; outgoing to the left be $b_i$; and outgoing to the 
right be $c_i$. The index $i$ ranges from $1$ to $N(\omega)$.
\beq
\begin{array}{c}\stackrel{a_i e^{i(\omega t - k_i x)}}{\rightarrow}\\ 
\stackrel{b_i e^{i(\omega t + k_i x)}}{\leftarrow}\\
\end{array}
\qquad
\begin{array}{c}\stackrel{c_i e^{i(\omega t-k_i x)}}{\rightarrow}\\ \stackrel{d_i e^{i(\omega t +k_i x)}}{\leftarrow}\\
\end{array}
\eeq
Then the S-matrix maps the incoming amplitudes to the outgoing amplitudes
\beq
S
\left( \begin{array}{c} \vdots\\ a_i\\ \vdots \\\vdots  \\ d_i\\ \vdots \\ \end{array} \right) = 
\left( \begin{array}{c} \vdots\\ b_i\\ \vdots \\\vdots  \\ c_i\\ \vdots \\ \end{array} \right) 
\eeq
The matrix elements are given by 
\beq
S = 
\left(\begin{array}{cc}
r & t'\\
t & r'\\
\end{array} \right)
\eeq
where $r$, $t'$, $t$ and $r'$ are all $N(\omega)\times N(\omega)$
matrices.
The matrix elements have the following interpretation:
$r_{ij}$ is the amplitude for a left moving wave in mode $i$ to be
reflected to mode $j$; ${r'}_{ij}$ is the amplitude for a right moving wave 
in mode $i$ to be
reflected into mode $j$; $t_{ij}$ is the amplitude for a left moving wave 
in mode $i$ to be into mode $j$;
and ${t'}_{ij}$ is the amplitude for a right moving wave in mode $i$ to be
transmitted into mode $j$.

For our purposes, the transfer matrix $R$ will prove more useful. 
It is defined by 
\beq
R
\left( \begin{array}{c} \vdots\\ a_i\\ \vdots \\\vdots  \\ b_i\\ \vdots \\ \end{array} \right) = 
\left( \begin{array}{c} \vdots\\ c_i\\ \vdots \\\vdots  \\ d_i\\ \vdots \\ \end{array} \right) 
\eeq
It has the property that if there are two regions in series labelled 
$0$ and $1$, then the  total transfer matrix is given by 
$R=  R_1 R_0$. It is not difficult to express $R$ in terms of the 
transmission and reflection amplitudes:
\beq
R = 
\left(\begin{array}{cc}
t-r't'^{-1}r & r't'^{-1}\\
-t'^{-1}r  & t'^{-1}\\
\end{array} \right)
\eeq
The total transmission coefficient is obtained by 
taking the $22$ element of $R_1 R_0$, and inverting it:
$t'_{0+1}= t'_0 ( 1-r_1 r'_0)^{-1} t'_1.$
This formula has a simple interpretation: write it as
$t'_{0+1}= t'_0t'_1 + t'_0 r_1r'_0 t'_1 +  t'_0 r_1r'_0 r_1r'_0  t'_1 +\cdots$.
It is evident that the total amplitude for transmission 
(from right to left) is given by
the amplitude to get through the second obstacle and then the first
obstacle; plus the amplitude to get through the second obstacle, 
bounce off the first obstacle, bounce off the second obstacle, 
and then get through the first, ad infinitum.

Thermal transport by vibrational modes in a multimode wire requires a few modifications to the above equations: starting from coupled linear equations of motion for many ionic coordinates, we derive the form of the dynamical transfer matrix $R$ that describes a longitudinal step along the wire.  As an example, first consider a purely one-dimensional chain of balls connected by springs of constant strength $k$ and mass $m$.  Let $u_i$ be the deviation, assumed small, of the position of ball $i$ from its equilibrium location.  The equation of motion for site $i$ in the chain is
\beq
m {\ddot u}_i = k (u_{i+1}-u_i) + k (u_{i-1}-u_i).
\eeq
The equation of motion at frequency $\omega$ can be written as follows:
\beq
\left(\matrix{u_{i+1}\cr u_i}\right) = \left(\matrix{2-{m \omega^2 / k}&-1\cr1&0}\right) \left(\matrix{u_i \cr u_{i-1}}\right).
\eeq
Hence the matrix that corresponds to adding a single site in this coordinate basis is
\beq
M = \left(\matrix{2-{m \omega^2 / k}&-1\cr1&0}\right).
\eeq
Now we convert this matrix to the basis of eigenmodes in order to obtain the transfer matrix $R$ defined above.

An eigenmode of wavevector $k_x$ satisfies (here $a$ is the lattice spacing)
\beq
M \left(\matrix{u_i \cr u_{i-1}}\right) = \lambda \left(\matrix{u_i \cr u_{i-1}}\right),\quad \lambda = e^{i k_x a}.
\eeq
$M$ is a real matrix and its eigenvalues form a time-reversed pair: they are either complex conjugates of each other (propagating modes in opposite directions) or reciprocal to each other (evanescent modes in opposite directions).  The eigenvalues are
\beq
\lambda = 1-{m \omega^2 \over 2 k} \pm {m \omega \sqrt{\omega^2 - 4 k /m } \over 2k}
\eeq
describing propagating modes with $k_x$ real if $m \omega^2 < 4 k$ and evanescent modes with $k_x$ imaginary if $m \omega^2 > 4.$  Consider the regime of propagating modes: in this basis the transfer matrix becomes just
\beq
R = \left(\matrix{\lambda &0 \cr 0&\lambda^*} \right),
\eeq
which gives a unitary $S$-matrix since
\beq
S = \left(\matrix{0 & \lambda \cr \lambda^* &0} \right)
\eeq and $|\lambda| = 1$.

The same procedure applies in a multimode wire except that now $r, t, r', t'$ are matrices.  There is one technical difference that appears between the vibrational case and the standard electronic case.  In the vibrational case, it is the energy flux rather than the probability flux that satisfies a continuity equation.  Hence the basis in which the $S$ matrix is (almost) unitary is one in which incoming and outgoing excitations are normalized to unit energy flux, rather than unit probability flux.  Actually the whole $S$ matrix is not unitary, but only the portion made up of modes with nonzero energy flux
\cite{transfootnote}.

Note that the transfer matrix approach used in this paper to make contact with mesoscopic electron physics works at a fixed frequency $\omega$ (equivalent to fixed energy in the electronic case).  The extension of the above to an extended 1D wire with random masses or spring constants is as follows: first the eigenmodes in the absence of disorder are obtained from the transfer matrix above.  Then the transmission coefficients for these modes through a disordered system are obtained by combining many individual segments (with different transfer matrices).  After the transmission coefficients for a fixed disorder realization are obtained as a function of $\omega$, a final integration step gives the thermal conductance for that realization as a function of temperature.  (Note that the first step of this approach, obtaining the eigenmode spectrum without disorder, is different computationally from typical dynamical matrix calculations in which $k_x$ is fixed and the frequencies $\omega$ of different phonon bands are obtained but obtains the same spectrum of propagating modes.)

In the multimode case, the transfer matrix $R$ describes how incident and outgoing modes at the left edge of a region are matched to incident and outgoing modes at the right edge of the region.  Once there is disorder in the system, energy flux incident in mode $i$ will be distributed over reflected modes, other transmitted modes, and the original mode, with conservation of energy flux.  The Landauer expression for the thermal conductance in Section II contains a sum over ${\cal T}_i$; ${\cal T}_i$ is defined as the fraction of the incident energy flux in mode $i$ that reaches the other end of the sample.
In the next section we show numerical results on the thermal conductance of 
two-dimensional disordered strips.


\section{Thermal conductance of quasi-one-dimensional systems: numerical results}
\label{sec:numerics}
\subsection{Two-dimensional strip}

This section applies the transfer matrix approach outlined in the previous section 
to a simple harmonic ``ball-and-spring'' model system with ions moving in two 
dimensions.  Considerable previous work~\cite{mahan} has shown in the absence of 
disorder that similar models, possibly including next-neighbor and bond-angle 
terms, can describe the phonon spectra of quasi-one-dimensional systems (including 
carbon nanotubes) with ions moving in three dimensions.  The justification for 
studying a model system with only two-dimensional motion is that already in two 
dimensions both linearly and quadratically dispersing modes are present, as in 3D, 
and the analytic predictions in section II can be checked comprehensively by going 
to very large wire sizes.  The following section (Section \ref{3Dsect}) sets out 
experimental predictions for semiconductor nanowires in 3D based on the analytic 
picture outlined in Section II and checked below.

The existence of gapless modes in a vibrational system is related to the Euclidean 
motions of a rigid body: the gapless vibrational modes are those that become 
symmetries, typically translations, in the long-wavelength limit.  There are four 
gapless modes in a rod, rather than three as in a bulk solid, because rotation around 
the rod axis guarantees a gapless ``torsional'' mode in addition to flexural 
and compressional modes that become translations along the three axes.  For a 
rigid wire moving in two dimensions, there are two gapless modes: using ${\hat x}$ 
to denote the rod axis, there is a linearly dispersing longitudinal mode with 
displacements along ${\hat x}$ in the long-wavelength limit, and a quadratically 
dispersing flexural mode with displacements along ${\hat y}$.  Additional modes 
begin to propagate as the frequency is increased.

\begin{figure}
\includegraphics[width=3in,height=1.5in]{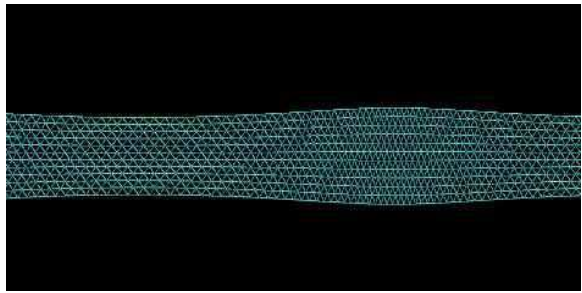}
\caption{The acoustic longitudinal mode of a strip of triangular
lattice.} 
\label{long_strip}
\end{figure}

\begin{figure}
\includegraphics[width=3in,height=1.5in]{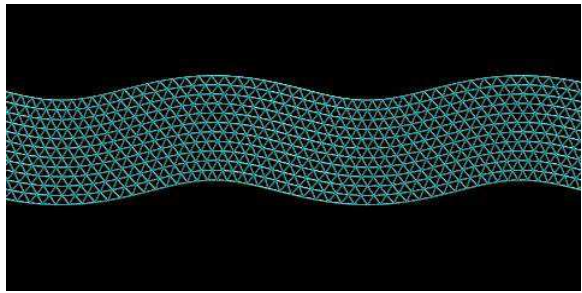}
\caption{The acoustic flexural mode of a strip of triangular
lattice.} 
\label{flex_strip}
\end{figure}

The simplest rigid body in two dimensions is the triangular lattice with springs 
connecting lattice points.  (The square lattice has vanishing rigidity in 2D if 
the potentials are purely length-dependent nearest-neighbor springs.)
The longitudinal mode for this system is shown in figure \ref{long_strip}
and the flexural mode is shown in figure \ref{flex_strip}.

A simple model for the number of propagating modes agrees 
qualitatively
with the numerical results for this quantity. 
The system has two modes as $\omega\rightarrow 0$ --- the longitudinal
and flexural acoustic modes.
As we increase $\omega$ to $\sim\pi v /2d$, where $d$ is the width, we
expect to encounter
modes that have a finite wavelength in the transverse direction. 
If we assume, in a simplified picture, that at each $\omega = n\pi v/2d$
we add 2 more modes to $N(\omega)$, and that every mode has the same velocity,
then the number of modes is approximately given by 
\beq
N(\omega) = 2 + {4 d \over \pi v } \omega
\eeq
For $\omega>\omega_D$, where $\omega_D$
is the Debye frequency, the lowest lying modes are no longer propagating, and
so one expects the number of modes to decrease, with approximately the
same slope.

In what follows we will assume that $N(\omega)$ has the form
\beq
N(\omega) = 2 + c { d \omega \over \omega_D}
\eeq
where $c$ is a constant of order unity, and $d$, the width, is
now measured in units of the lattice spacing.

\begin{figure}
\includegraphics[scale=0.65]{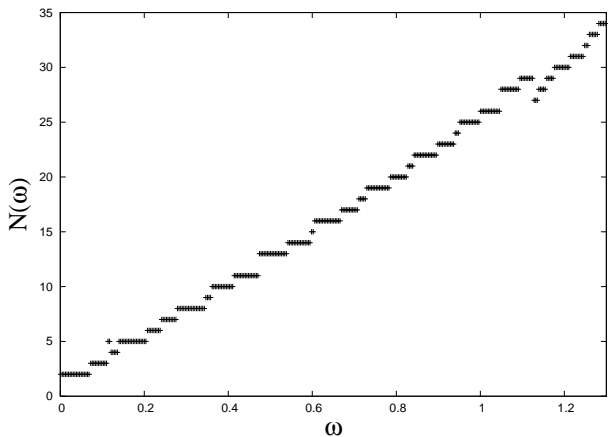}
\caption{The number of modes $N(\omega)$ for a strip of width 32 sites for $\omega<\omega_D$. $N(\omega)$ is approximately linear in $\omega$ for this 2D system, justifying
the form for $N(\omega)$ used in the text.} 
\label{number_of_modes}
\end{figure}

\subsubsection{Bulk disordered}

Consider the case where the isotopic disorder extends 
throughout the material (in a later section we will
constrain the disorder to be at the edge of the strip).
The rate at which waves scatter from point-like defects 
is determined by the Rayleigh-Klemens formula ~\cite{klemens}:
\beq
\tau(\omega)^{-1} = \sigma^2 \omega^2 \left( {\cal D} {\omega^{{\cal D}-1}\over 
\omega_{D}^{\cal D}}\right)
\eeq
where ${\cal D}$ refers here to the dimension of the space, and $\sigma^2$
is the fluctuation in a single mass.
The DMPK formalism suggests a formula for the total throughput
of a quasi-one-dimensional system in the ballistic and Ohmic
regimes:
\beq
{\cal T}(\omega) = {N(\omega) \over 1 + L/\ell(\omega)}
\eeq
As the length of the system $L$ extends belong the mean free
path $\ell$, the throughput (and hence the contribution
of that frequency to the thermal conductance) scales like 
$1/L$; this is a feature characteristic of Ohmic systems.

From the Rayleigh-Klemens formula, one anticipates a mean free path
of
\beq
{1\over\ell(\omega)} \sim {1\over v} \sigma^2 \omega^2 N(\omega)
\eeq
In the Ohmic limit this suggests a length and frequency scaling
in the throughput of
\beq
{\cal T}(\omega) \simeq {N(\omega) \ell(\omega) \over L} 
\sim {N(\omega) \over L} {1\over \omega^2 N(\omega)}
\sim {1\over L \omega^2}
\eeq

\begin{figure} 
\includegraphics[scale=0.65]{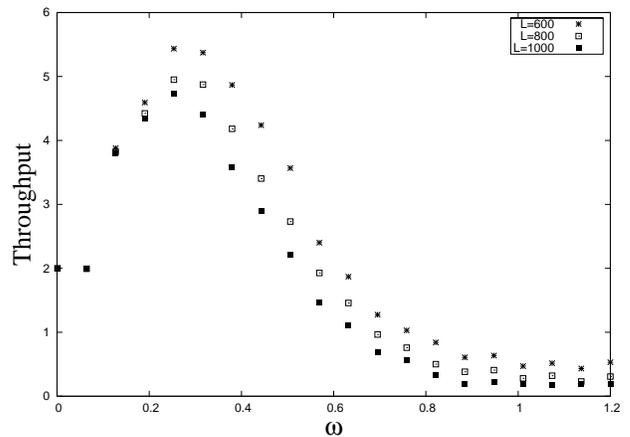}
\caption{The throughput as a function of frequency for a two-dimensional strip
of width 32 sites. The various lengths are shown in the legend. The throughput
shown has been averaged over 40 implementations of the disorder, as described in
the text.
} \label{bulk5}
\end{figure}

\begin{figure} 
\includegraphics[scale=0.65]{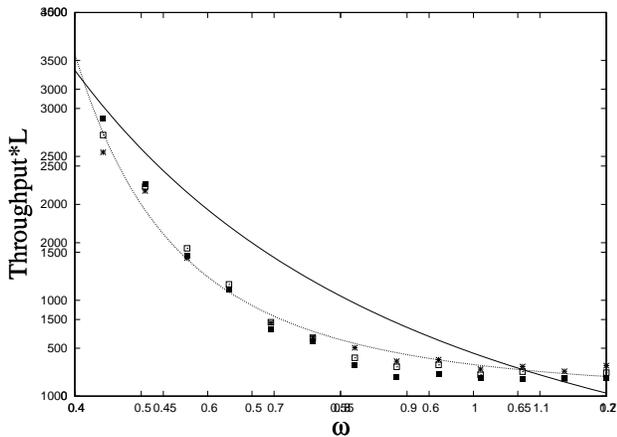}
\caption{This figure shows the same data as figure \ref{bulk5}; 
here though the throughput has been multiplied by the length 
of the system, to make the $1/L$ dependence more clear. 
The curve shown is a fit to $A/\omega^{2.6}$.} 
\label{bulk4}
\end{figure}

As an example of the throughput of a bulk disordered system, see
figures \ref{bulk5}, \ref{bulk4}. 
The strip chosen was of width 32 sites, and of varying lengths 
as shown in the legend. In each case, the probability that a given
site was replaced by an disordered site is $1/32$. Here the
disordered sites were chosen to have mass $1+\sqrt{2}$, whereas 
the original masses are $1$.

The general shape of the curves in figure \ref{bulk5} is consistent with
the DMPK formula for the throughput, with the frequency
dependent mean free path given by the Rayleigh-Klemens
formula. For small frequencies, the phonon scattering rate is
small, and so the total throughput tracks the number of modes.
As the frequency increases, the phonons are more strongly scattered;
the mean free path decreases faster than the density of states increases, 
and so the throughput decreases as with increasing frequency.

The dependence of the throughput on $L$ and $\omega$ is considered
in figure \ref{bulk4}. In this figure the throughput multiplied
by the length is plotted as a function of frequency.
The throughput clearly scales like $1/L$ in the frequency regime shown. 
The line shown in the figure is of the form $A/\omega^{\alpha}$, where
$A$ and $\alpha$ are fit to the data. 

\begin{figure}
\includegraphics[scale=0.65]{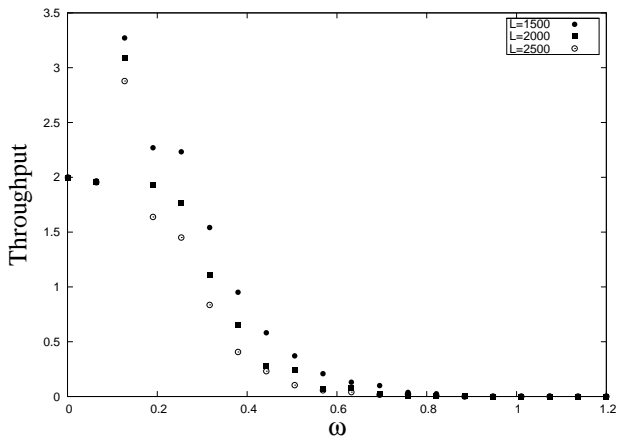}
\caption{Throughput as a function of frequency for a bulk disordered strip 
of width 32 sites and various lengths. The alloy fraction here is 0.1. As the throughput
passes through $1$, it tends to zero faster than any power. 
}
\label{loc}
\end{figure}

The throughput for the localized regime is illustrated in figure 
\ref{loc}. As the Ohmic throughput, ${\cal T} \sim N\ell/L$ 
decreases to a value less than $1$, the throughput decreases faster
than any power law. 

\subsubsection{Edge disordered}

We consider in this section the throughput of the two dimensional
strip with only the boundary of the strip disordered. 
Recall that 
in the case of bulk disorder, the throughput shows a strong frequency
dependence --- essentially because the mean free path
is set by Rayleigh-Klemens formula for scattering from point-like
defects. In contrast, when the disorder is present only at the edge, 
a more appropriate model is that the phonons are scattered, specularly or diffusively,
at the boundary.

\begin{figure}
\includegraphics[scale=0.65]{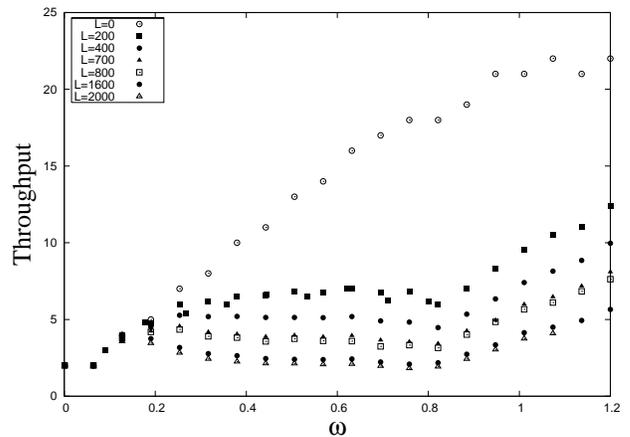}
\caption{Throughput as a function of frequency. The strip is edge-disordered 
and of width 32 sites.}
\label{edge}
\end{figure}

\begin{figure}
\includegraphics[scale=0.65]{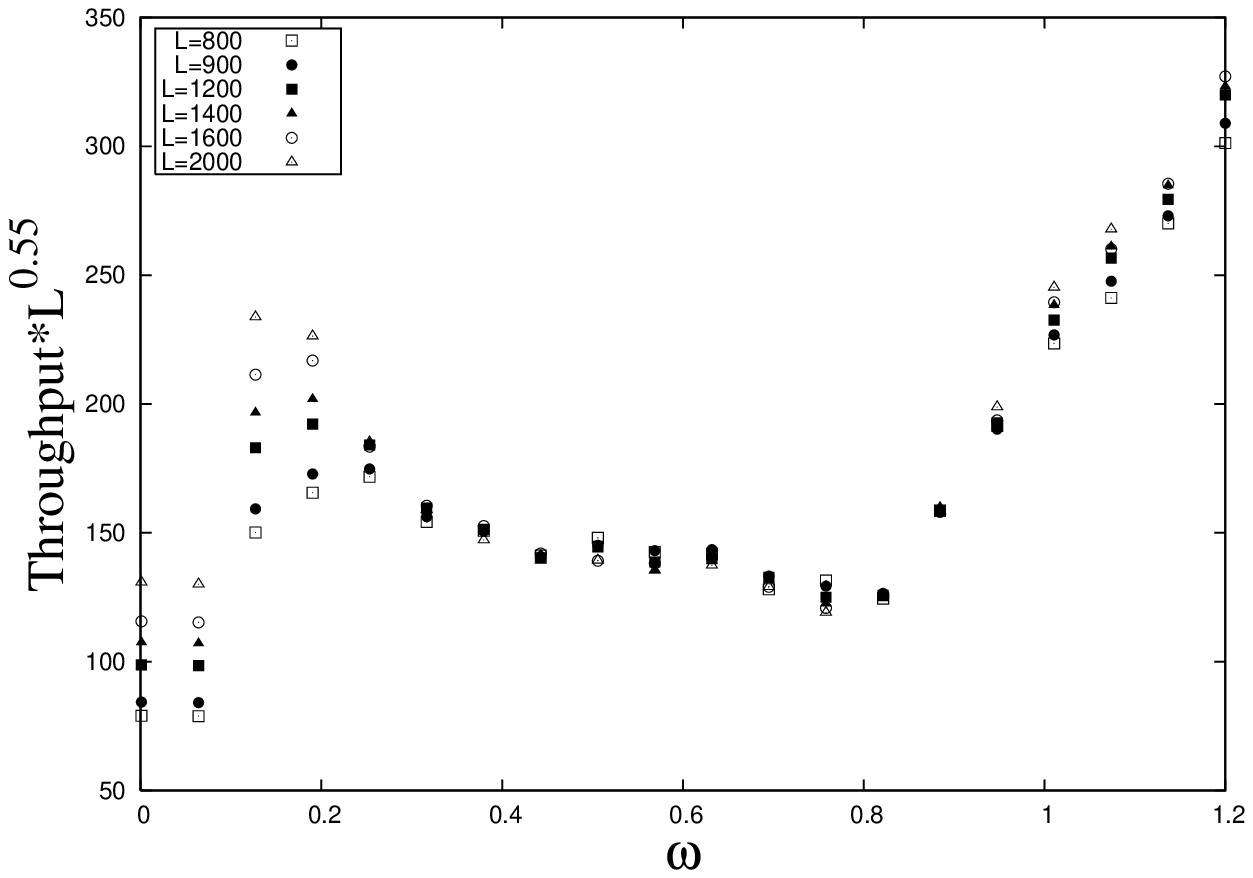}
\caption{Throughput of a strip of width 32 sites multiplied by $L^{0.55}$
as a function of frequency for the case where the disordered is
only at the edge.}
\label{edge4}
\end{figure}

\begin{figure}
\includegraphics[scale=0.65]{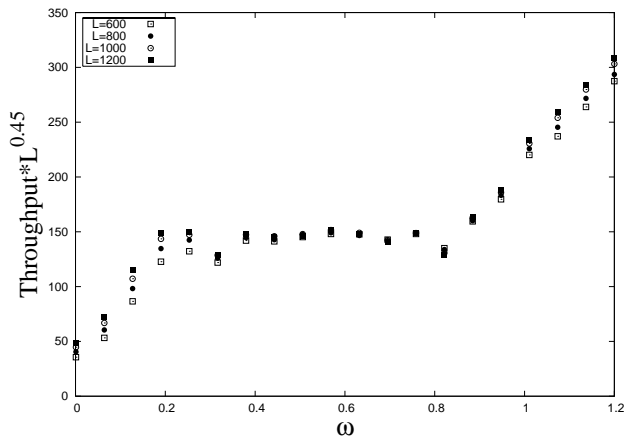}
\caption{Throughput of a strip of width 48 sites multiplied by $L^{0.45}$
as a function of frequency for the case where the disordered is
only at the edge.}
\label{edge4_48}
\end{figure}

The throughput is shown as a function of frequency for various 
lengths in figures \ref{edge}, \ref{edge4}, and \ref{edge4_48}. 
The width of the strip in figures \ref{edge} and \ref{edge4}
 is 32 sites, and in figure \ref{edge4_48} is 48. The throughput was calculated by
disordering a region of thickness 1 site at both edges of the strip. The 
mass of these edge sites was chosen at random to be either 1 (the mass
in the clean case) or $1+\sqrt{2}$, each with probability 1/2. 
The throughput was then calculated numerically as described earlier. 
The figures show the throughput averaged over 40 implementations
of the disorder.

From the figures one can see that 
once the phonon frequency is no longer in the ballistic
regime, the throughput levels off to approximately a constant.
A throughput that is constant as a function of frequency 
can be seen experimentally as a thermal conductance that is
linear in temperature: if the throughput is equal to say $c$, then
the thermal conductance is $c g_0$, where $g_0$ is the quantum 
of thermal conductance. The throughput has been rescaled to
remove the length dependence. In both cases the throughput scales like
$1/L^\alpha$, where $\alpha$ is less than 1: the system is not in the 
Ohmic regime.

\begin{figure}
\includegraphics[scale=0.65]{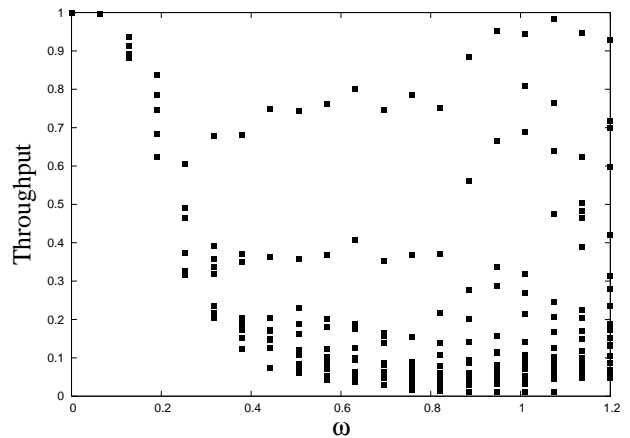}
\caption{
Mode-resolved throughput of a strip of width 32 sites 
as a function of frequency for the case where the disordered is
distributed only at the edge of the material. The length of the strip
is 1600 sites.}
\label{mm_edge}
\end{figure}

To illustrate the reason for the frequency independent throughput, 
we show the mode-resolved throughput in figure \ref{mm_edge}.  In the frequency range
in which it is constant, the throughput is dominated by two modes, 
which happen to have a small displacement at the boundary. 
Their contribution is approximately constant.


\begin{figure}
\includegraphics[scale=0.65]{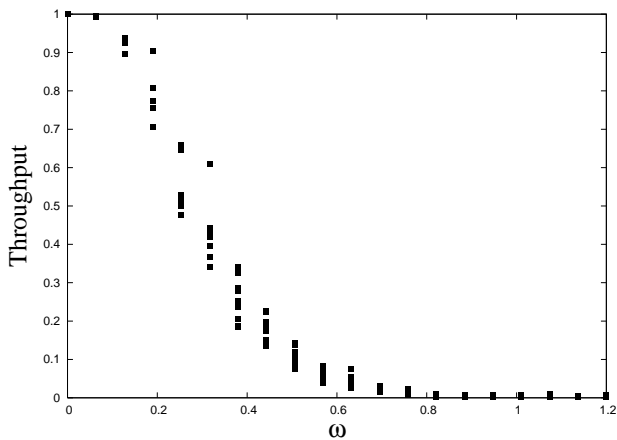}
\caption{
Mode-resolved throughput of a strip of width 32 sites 
as a function of frequency for the case where the disordered is
distributed throughout the bulk of the material. The length of
the strip is 1600 sites.}
\label{mm_bulk}
\end{figure}

When the disorder is exclusively at the boundary, the boundary
displacement of the different modes will obviously determine 
how strongly they are scattered --- a mode with vanishing boundary
displacement will pass through the material ballistically. 

In addition, these boundary displacements can vary dramatically
for different modes at the same frequency. For instance, when
$\omega$ is such that $\omega > v/d$, the ungapped longitudinal
and flexural modes become surface waves. This occurs because 
a surface wave must decay exponentially into the bulk. 
Its dispersion must then be $\omega \sim v\sqrt{k_z^2 - k_x^2}$. 
Therefore for a given $\omega$, these surface modes have the largest $k_z$ 
among all the propagating modes: this identifies them as a linear combination 
of the flexural and longitudinal modes.

By contrast, the first set of gapped modes tend to have an anomalously
small displacement at the boundary. This is illustrated in table \ref{table:displac}. 
For a frequency of $\omega = 0.57$, and a width of 32 sites, the 
$k_z$ values of the various propagating modes along with the amount of
displacement at the boundary is shown. By boundary displacement
we mean the displacement summed over the two sites at the edge of the strip.
The modes here are normalized to 
$1/2$: a mode with boundary displacement squared $1/2$ would exist exclusively
at the top and bottom sites of the strip.

\begin{table}
\begin{center}
\begin{tabular}{|r|r|}
\hline
 & \\
$k_{xj}$ & $|\vec\phi|^2$ \\
 &  \\
\hline
1.059383 &      0.189611  \\
1.059383 &      0.189612  \\
0.929791 &      0.002395  \\
0.904105 &      0.008771  \\
0.854861 &      0.016750  \\
0.773922 &      0.024001  \\
0.657264 &      0.033273  \\
0.558262 &      0.088231  \\
0.526373 &      0.087008  \\
0.525994 &      0.025183  \\
0.447062 &      0.076678  \\
0.442868 &      0.038433  \\
0.275598 &      0.062901  \\
0.293598 &      0.045720  \\
\hline
\end{tabular}
\caption{\label{tab:bdry_displac} Boundary displacements for the
14 propagating modes at frequency $\omega = 0.57$ for a strip of width 
32 sites.}
\label{table:displac}
\end{center}
\end{table}

\subsection{Three-dimensional rod}\label{3Dsect}

\begin{figure}
\includegraphics[scale=0.25]{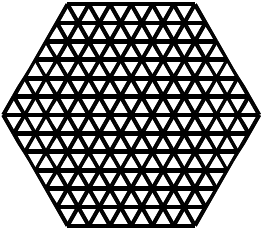}
\caption{A cross-section of the nanowires considered. There
are in fact springs connecting next-to-nearest neighbors as
well as nearest neighbors, although they are not shown here.
} \label{xsection}
\end{figure}

In this section we consider numerical data for a three-dimensional
rod. The system we study has 127 atoms in a cross-section, as shown in figure
\ref{xsection}.
The atoms are arranged in a triangular lattice, with 
springs connecting nearest neighbors and next-to-nearest neighbors.
At small frequencies there are four normal modes: longitudinal, 
torsional and two flexural. The longitudinal and torsional
modes disperse linearly, while the flexural disperse
quadratically. 

The rods are disordered in the same way as described for the
strip. In the clean case each atom has mass 1; for a disordered
system each atom is replaced by an atom of mass $1+\sqrt{2}$
with a certain probability, the alloy fraction. As before, we will consider 
both bulk (i.e., each atom in a given cross-section can be replaced)
and boundary (i.e., only those atoms at the outer edge of the 
cross-section can be replaced) disorder.

\begin{figure}
\includegraphics[scale=0.65]{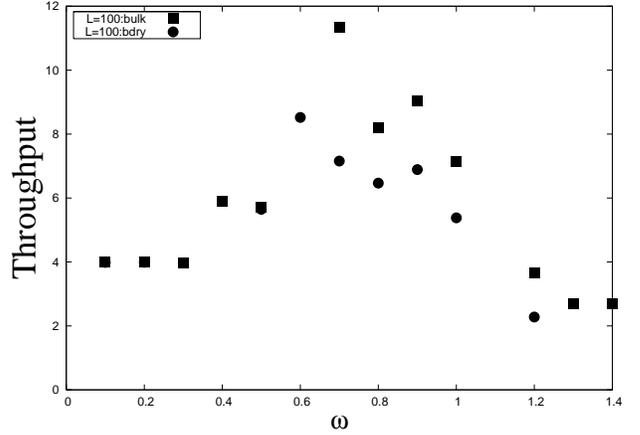}
\caption{ 
Disorder-averaged throughput for a rod of length of length 100 sites as 
a function of frequency. Both the bulk and boundary disordered cases are shown.
}\label{tpR7}
\end{figure}

We find the throughput using the transfer matrix method. In figure \ref{tpR7}, 
we show the disorder-averaged throughput for a rod of length 100 sites.
Both bulk and boundary disordered cases are shown. The alloy fraction 
for the bulk case was chosen to be 0.1; for the boundary case it was
chosen to be 12.7/36: this ensures that the number of disordered atoms
in a cross-section is the same for the bulk and boundary disordered cases. 

\begin{figure}
\includegraphics[scale=0.65]{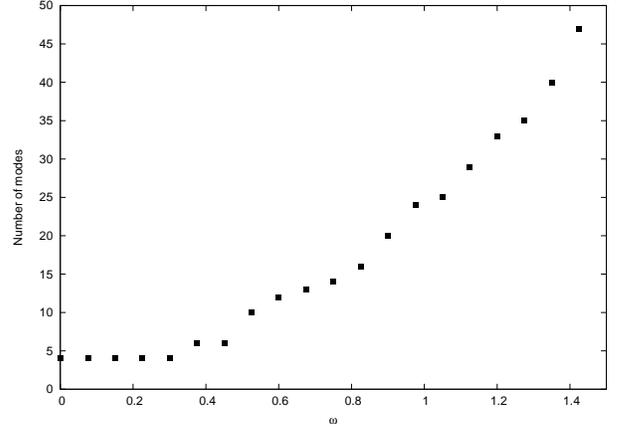}
\caption{ The number of propagating modes for the rod studied as a
function of frequency.
} \label{count_modes_rod}
\end{figure}

The data shows that the throughput in the case of boundary
scattered rods is noticeably less than the bulk-disordered case. 
This is because phonon modes in a rod tend
to have more displacement at the edge than at the center, 
since at the edge the masses have greater freedom to vibrate.\cite{pochhammer}

One feature which is not apparent is the presence
of modes which have a displacement at the boundary
orders of magnitude smaller than that of the other modes
(recall that these modes dominated the throughput of the edge-disordered 
strip). This may be due to the relatively small diameter of the
rod considered here: in the data shown, we have not in fact reached
the limit where the surface waves have formed.

Finally, it is interesting to note that the reflection and 
transmission matrices for a short system can be used to find
the incoherent limiting conductance.
If  the transmission
and reflection matrices for a system of length $\delta L$ 
are $\tilde{ T}$ and $\tilde{R}$, then
the transmission and reflection matrices for a 
system of length $n\delta L$ (denoted $T_n$ and $R_n$) are
\beq
\label{eq:trans}
T_n = \tilde{T} ( 1 - R_{n-1} \tilde{R} )^{-1} T_{n-1}\\
\eeq
\beq
\label{eq:refl}
R_n = R_{n-1} + T_{n-1} \tilde{R} ( 1 - R_{n-1} \tilde{R} )^{-1} T_{n-1}
\eeq
This is of course an approximate description of the
transport of waves: it is analagous to adding
the probabilities rather than amplitudes, and so it
neglects interference effects. However, a physical system
will have a phase-breaking length scale; for systems
longer than this, the transport will be incoherent. 
Physically one would expect this phase-breaking scale to depend on
frequency and temperature.  This may prove useful to isolating the 
incoherent signal in future studies.

\section{Conclusion}

In conclusion, we have studied the thermal
conductance of insulating nanowires at low temperatures. 
For the thinnest wires, with diameters of order $20$ nm, it was argued that
frequency-dependent surface scattering is relevant. 
In particular, modes with a small $k_\perp$ (ie.~the component of the
wavevector in the direction perpendicular to the long direction of the wire)
should scatter almost specularly from the boundary, whereas those with a larger 
$k_\perp$ presumably have a mean free path of order $d$, the width of the wire.
The resulting thermal conductance is shown in figure \ref{thermal_cond_num_int}, 
for a number of diameters. For a width $d=20$ nm, the thermal conductance is 
almost equal to that of a single ballistic mode. This can be understood as
follows. There are two contributions to the thermal conductance: that of
the quasi-specular modes, whose number is fixed but whose contribution
decreases with increasing frequency; and the diffusively scattered modes, 
whose mean free path is fixed but whose number increases with frequency. 
These decreasing and increasing terms together give a approximately 
constant throughput for low frequencies (figure \ref{tp_analytic}).

These considerations also suggest strongly that localization effects
are relevant to transport in nanorods of transverse dimension $d \lesssim 20$ nm. 
This was demonstrated by applying the DMPK formulism, which at certain
frequencies is not valid --- DMPK requires mode equivalence, and at
small frequencies the quasi-specular mode has a mean free path 
much longer than the other modes. Nevertheless, it would be of great
interest to explore experimentally localization phenomena in these nanowires. 
For instance, a different temperature dependence in the thermal
conductance and heat capacity, with the heat capacity measured through a bulk
contact to a reservoir, would be strong evidence that some of the phonon
modes are localized.

In order to understand
the relevance of coherence effects,
we have also simulated phonon transport using a dynamical transfer
matrix method. 
Our simulations show a mode inequivalence 
for the case of scattering from a disordered surface that is 
not present when the impurities are placed at random throughout the strip. 
The total throughput of the surface-disordered strip is approximately
constant in frequency, consistent with a linear thermal conductance.

If the ballistic and strongly scattered modes were decoupled, one would
expect some modes to have a throughput of approximately $1$, and others to
have an exponentially small throughput. 
Although the mode inequivalence we observe numerically is not that strong,
the total throughput is
consistent with the picture that some modes are quasi-ballistic, and the rest 
are strongly scattered.  Also, the throughput for the surface scattering case shows a 
non-Ohmic behavior approximately described by a power-law: ${\cal T}(\omega) \sim 1/L^{\alpha}$ with $\alpha <1$.
This is consistent with the presence of quasi-ballistic modes. 

An additional effect brought to light by our simulations is the
highly variable displacement that phonon modes can have at the boundary of
the nanowire. We found that whereas the gapless modes acquire a 
large boundary displacement, other modes acquire anomalously low
values. These were the modes that gave a quasi-ballistic
contribution to transport.  Our dynamical transfer matrix method could
be applied to specific crystal structures and materials of interest
by using parameters from full ab initio simulations of nanowires, including
boundary effects.

\acknowledgments

The authors wish to thank Renkun Chen, Allon Hochbaum, Arun Majumdar, and Peidong Yang for useful conversations.  This work was supported by the DOE-BES Thermoelectrics Program at Lawrence Berkeley National Laboratory.

\bibliographystyle{./apsrev}
\bibliography{./bigbib_nanowire}

\begin{thebibliography}{29}
\bibitem[{tra()}]{transfootnote}
\bibinfo{note}{If $\lambda_i$ are the amplitudes in orthogonal incident modes
  $i$ with energy flux $v_i$, while $\lambda_o$ are the amplitudes of
  orthogonal outgoing modes $o$ with flux $v_o$, then the flux conservation
  condition is \beq \sum_i |\lambda_i|^2 v_i = |\lambda_o|^2 v_o \eeq which
  becomes equivalent to a unitarity condition if all modes have nonzero flux.
  In general, evanescent modes are required to make a complete eigenbasis but
  have zero energy flux; in this case the requirement of flux conservation does
  not affect elements of the $S$ matrix that involve evanescent modes. For
  example, for $m \omega^2 > 4$ in the above example the $S$ matrix is
  nonunitary in the eigenmode basis but energy flux is still conserved.}
\end{thebibliography}

\end{document}